\documentclass[a4paper,USenglish]{lipics}

\usepackage{color}
\usepackage{amsmath,amssymb}
\usepackage{graphicx}
\usepackage{tikz}
\usetikzlibrary{patterns,arrows,shapes,decorations.pathreplacing}
\usepackage[utf8]{inputenc}
\usepackage{times}
\usepackage{verbatim}
\usepackage{bbm}
\usepackage{mathtools}
\usepackage{microtype}
\usepackage{enumerate}
\usepackage{url}

\title{Balanced Interval Coloring}

\author{Antonios Antoniadis}
\author{Falk Hüffner}
\author{Pascal Lenzner}
\author{Carsten Moldenhauer}
\author{Alexander Souza}

\authorrunning{A. Antoniadis et al.}

\affil{%
  Institut für Informatik, Humboldt-Universität zu Berlin,
  D-10099 Berlin, Germany
  \texttt{\{antoniad,hueffner,lenzner,moldenha,souza\}@informatik.hu-berlin.de}
}

\Copyright[nd]{Antonios Antoniadis and Falk Hüffner and Pascal Lenzner and Carsten Moldenhauer and Alexander Souza}


\newcommand{\N}{\mathbbm{N}}
\newcommand{\R}{\mathbbm{R}}

\newcommand{\bic}[1]{{\sc Minimum Imbalance Interval $#1$-Coloring}}
\newcommand{\bbc}[2]{{\sc Minimum Imbalance $#1$-Box $#2$-Coloring}}
\newcommand{\dbbc}[2]{{\sc Balanced $#1$-Box $#2$-Coloring}}
\newcommand{\kbic}{\bic{k}}
\newcommand{\dkbbc}{\bbc{d}{k}}
\newcommand{\ibbc}{\dbbc{d}{k}}

\newcommand{\col}{c}

\DeclareMathOperator{\imb}{imb}
\DeclareMathOperator{\simb}{simb}
\DeclareMathOperator{\dsc}{disc}

\begin{document}

\maketitle

\begin{abstract}
We consider the discrepancy problem of coloring $n$~intervals with
$k$~colors such that at each point on the line, the maximal difference
between the number of intervals of any two colors is minimal. Somewhat
surprisingly, a coloring with maximal difference at most one always
exists. Furthermore, we give an algorithm with running time $O(n \log
n + kn \log k)$ for its construction. This is in particular
interesting because many known results for discrepancy problems are
non-constructive. This problem naturally models a load balancing
scenario, where $n$~tasks with given start- and endtimes have to be
distributed among $k$~servers. Our results imply that this can be done
ideally balanced.

When generalizing to $d$-dimensional boxes (instead of intervals), a
solution with difference at most one is not always possible. We show
that for any $d \ge 2$ and any $k \ge 2$ it is NP-complete to decide if such a solution exists, which implies also NP-hardness of the respective minimization
problem.

In an online scenario, where intervals arrive over time and
the color has to be decided upon arrival, the maximal difference in
the size of color classes can become arbitrarily high for any online
algorithm.
\end{abstract}

\section{Introduction}

In this paper, we consider the following load balancing problem: We
are given a set $\mathcal{I} = \{ I_1, \dots, I_n \}$ of tasks, where
each task is represented by an interval $I = [\ell, r] \in
\mathcal{I}$ with starttime~$\ell$ and endtime $r$. Furthermore, we
are given $k$ servers and have to assign the tasks to the servers as
evenly as possible. That is, we want to minimize the maximal
difference of the numbers of tasks processed by any two servers over
all times.


We formalize this in terms of an interval coloring problem: We are given a set $\mathcal{I} = \{ I_1, \dots, I_n \}$ of $n$ intervals on the real line and a set $K = \{1, \dots, k\}$ of $k$ colors. A \emph{$k$-coloring} is a mapping $\chi : \mathcal{I} \rightarrow K$. For a fixed $k$-coloring $\chi$ and a point $x \in \R$, let $\col_i(x)$ denote the number of intervals containing $x$ that have color $i$ in $\chi$. Define the \emph{imbalance} of $\chi$ at $x$ by
\begin{align}
\imb(x) = \max_{i, j \in K} | \col_i(x) - \col_j(x) |.
\end{align}
In words, this is the maximum difference in the size of color classes at point $x$. The \emph{imbalance} of $\chi$ is given by
$\imb(\chi) = \max_{x \in \R} \imb(x)$.

These definitions yield the following minimization problem:

\begin{center}%
  \begin{minipage}{0.9\linewidth}%
	\kbic\\
	\textbf{Instance:} A set of intervals $\mathcal I$.\\
	\textbf{Task:} Find a $k$-coloring $\chi$ with minimal $\imb(\chi)$.
  \end{minipage}%
\end{center}%

We call a $k$-coloring with imbalance at most one
\emph{balanced}. Observe that if the number of intervals intersecting
at some point is not divisible by $k$, then imbalance at least one is
unavoidable. On the other hand, if the number of intersecting
intervals is divisible by~$k$, then no coloring having imbalance one
exists. Thus, if a
balanced coloring exists, its imbalance is minimal. 

As we
will see shortly, it is always possible to find a balanced interval
$k$-coloring. Hence, we will mostly be concerned with its
construction. More specifically, the questions considered in this
paper are outlined as follows:

\begin{enumerate}[(i)]
\item Is there always a balanced $k$-coloring? \label{it:q1}
\item If so, is it possible to construct a balanced $k$-coloring in polynomial time? \label{it:q2}
\item If we consider arcs of a circle (instead of intervals), do balanced $k$-colorings always exist? \label{it:q3}
\item How is the situation if intervals arrive online? \label{it:q4}
\item If $d$-dimensional boxes (instead of intervals) are considered, can the existence of a balanced $k$-coloring be decided in polynomial time? \label{it:q5}
\end{enumerate}

The problem has close connections to discrepancy theory; see
Doerr~\cite{Doerr05} and Matoušek~\cite{Mat99} for introductions to
the field.
Let $H = (X, U)$ be a hypergraph consisting of a set $X$ of vertices and a set $U \subseteq 2^X$ of hyperedges. Analogous to the previous definitions, a $k$-coloring is a mapping $\chi : X \rightarrow K$, and the imbalance $\imb(\chi)$ is the largest difference in size between two color classes over all hyperedges. The \emph{discrepancy} problem is to determine the smallest possible imbalance, i.\,e., $\dsc(H) = \min_{\chi : X \rightarrow K} \imb(\chi)$.

Hence our problem is to find the discrepancy of the hypergraph $H = (I, U)$, where $U$ is the family of all maximal subsets of intervals intersecting at some point. It turns out that this hypergraph has totally unimodular incidence matrix,
which is useful because de Werra~\cite{Wer71} proved that balanced
$k$-colorings exist for hypergraphs with totally unimodular incidence
matrix. However, the proof in~\cite{Wer71} is only partially
constructive: A balanced $k$-coloring is constructed by iteratively
solving the problem of balanced $2$-coloring on hypergraphs with
totally unimodular incidence matrix, for which no algorithm was given
in~\cite{Wer71}.

Further related work in discrepancy theory mostly considers hypergraph coloring with two colors and often from existential, rather than algorithmic perspective. For an arbitrary hypergraph $H$ with $n$ vertices and $m$ hyperedges, the bound $\dsc(H) \le \sqrt{2n \ln(2m)}$ for $2$-coloring follows with the probabilistic method; see also~\cite{Doerr05}. For $m \ge n$, Spencer~\cite{Spe85} proved the stronger result $\dsc(H) = O(\sqrt{n \log(m/n)})$, which is in particular interesting for $m = O(n)$. If each vertex is contained in at most $t$ edges, the $2$-coloring bound $\dsc(H) = O(\sqrt{t} \log n)$ was shown by Srinivasan~\cite{Sri97} and the bound $\dsc(H) \le 2t-1$ by Beck and Fiala~\cite{BF81}. Biedl et al.~\cite{BCC02} improved the bound to $\dsc(H) \le \max\{2t-3, 2\}$ for $2$-colorings and established $\dsc(H) \le 4t-3$ for general $k$-colorings. They also showed that it is NP-complete to decide the existence of balanced $k$-colorings for hypergraphs with $t \ge \max\{3, k-1\}$ and $k \ge 2$.

Bansal~\cite{Ban10} recently gave efficient algorithms that achieve $2$-color imbalances similar to~\cite{Spe85,Sri97} up to constant factors. In particular, an algorithm yields $\dsc(H) = O(\sqrt{n} \log(2m / n))$ matching the result of Spencer~\cite{Spe85} if $m = O(n)$. Furthermore, $\dsc(H) = O(\sqrt{t} \log n)$ complies with the non-constructive result of Srinivasan~\cite{Sri97}. For general $k > 2$, Doerr and Srivastav~\cite{Doer03} gave a recursive method constructing $k$-colorings from (approximative) $2$-colorings.

Unfortunately, these results on general discrepancy theory do not
answer any of the problems considered here, because 
$t$ is only bounded by the number of vertices.

\subparagraph{Our Contributions.}
We contribute the following answers to the above questions:
\begin{enumerate}[(i)]
\item Balanced $k$-colorings exist for any set $\mathcal{I}$ of
  intervals, i.\,e., question~\eqref{it:q1} can \emph{always} be
  answered in the affirmative. We establish this by showing that our
  hypergraph $H$ has totally unimodular incidence matrix and then applying
  a result of de Werra~\cite{Wer71}. This also follows independently
  from our algorithmic results below.
\item We present an $O(n \log n)$ time algorithm for finding a balanced $2$-coloring,
  thereby establishing the first constructive result for
  intervals. 
  Furthermore, we give an $O(n \log n + kn \log
  k)$ algorithm for finding a balanced $k$-coloring. This is an improvement in
  time complexity, since the construction of de Werra~\cite{Wer71}
  combined with our algorithm for $2$-coloring only yields $O(n \log n + k^2
  n)$.  We also note that our algorithm works for any hypergraph with
  incidence matrix having the consecutive-ones property.
\item If we consider arcs of a circle instead of intervals, balanced
  $k$-colorings do not exist in general. However, we give an algorithm
  achieving imbalance at most two with the same time complexity as in
  the interval case.
\item In an online scenario, in which we learn intervals over time, the imbalance of \emph{any} online algorithm can be made arbitrarily high.
\item For $d$-dimensional boxes, it is NP-complete to decide if a balanced $k$-coloring exists for any $d \ge 2$ and any $k \ge 2$. Our reduction is from \textsc{Not-All-Equal 3SAT}. This result clearly implies NP-hardness of the respective minimization problem.
\end{enumerate}



\section{Interval Colorings}

In this section, we consider \kbic, establish the existence of balanced $k$-colorings, and give algorithms for $2$ and $k$ colors, respectively. Later, we consider arcs of a circle and an online version.

\subsection{Existence of Balanced $k$-Colorings}
\label{sec:ex}

We begin by observing the existence of balanced $k$-colorings. In the proof below, we use a theorem of de Werra~\cite{Wer71}, but the existence of balanced $k$-colorings also follows from our algorithmic results.

\begin{theorem}
\label{thm:existence}
For any set $\mathcal{I}$ of intervals and any $k \in \N$, there is a balanced $k$-coloring.
\end{theorem}

\begin{proof}
Let $\mathcal{I}$ be the set of given intervals. Define a hypergraph $H = (I, U)$, where $U$ is the family of all maximal subsets of intervals intersecting at some point. For $H$ with $I = \{I_1, \dots, I_n\}$ and $U = \{ U_1, \dots, U_m \}$, the incidence matrix is defined by $A = (a_{i,j})$ with $a_{i,j} = 1$ if $I_i \in U_j$ and $a_{i,j} = 0$ otherwise.

De Werra~\cite{Wer71} showed that any hypergraph with totally unimodular
incidence matrix admits a balanced $k$-coloring. It is well-known~\cite{Schrijver86}
that a $0$--$1$-matrix is totally unimodular if it has the
consecutive-ones property, i.\,e., if there is a permutation of its
columns such that all $1$-entries appear consecutively in every
row. The incidence matrix $A$ of $H$ has this property: If we order
the $U_j$ in increasing order of intersection points, then the entries
$a_{i,j} = 1$ appear consecutively in each row.
\end{proof}

\subsection{Algorithm for Two Colors}
\label{sec:k2}

In this section, we present an algorithm that constructs a balanced $2$-coloring
in polynomial time. Since the algorithm produces a valid solution for every possible
instance, Theorem~\ref{thm:existence} for $k=2$ also follows from this algorithmic result.
We note in passing that a polynomial-time algorithm can also be
obtained by solving a simple Integer Linear Program (ILP) with a
totally unimodular constraint matrix. However, this gives a much worse
running time bound of~$O(n^5/\log n)$~\cite{Ans99}.

The main idea of our algorithm is to simplify the structure of the instance such that
the remaining intervals have start- and endpoints occurring pairwise.
We then build a constraint graph that has the intervals as vertices.
Finally, a proper $2$-coloring of the constraint graph induces a solution
to the problem.

The start- and endpoints of the intervals are called \emph{events}.
A \emph{region} is an interval spanned by two consecutive events
and is called even (odd) if it is contained in an even (odd) number of
input intervals.

\begin{theorem}
	For any set $\mathcal I$ of $n$ intervals, there is a balanced $2$-coloring
	that can be constructed in $O(n \log n)$ time.
\end{theorem}
\begin{proof}
        W.\,l.\,o.\,g.\ we can assume that the start- and endpoints of the input intervals
	are pairwise disjoint. If not, a new instance can be obtained by repeatedly
	increasing one of the coinciding start- or endpoints by $\epsilon/2$,
	where $\epsilon$ is the minimum size of a region.
	Since the new instance includes a corresponding region for every region of the original
	instance, a balanced coloring for the new instance is a balanced coloring for the old instance
	(the converse is not true).

	Observe that a coloring that is balanced on all even regions is also balanced
	on all odd regions. This is because odd regions only differ by one interval
	from a neighboring even region.
	Thus, the task reduces to constructing a balanced coloring of the even regions.
	Since between two consecutive even regions, exactly two events occur,
	it suffices to consider only pairs of consecutive events enclosing odd regions.

	If a pair of events consists of the start- and endpoint of the same interval,
	this interval is assigned any color and is removed from the instance.
	If a pair consists of start- and endpoint of different intervals, these intervals
	are removed from the instance and substituted by a new (minimal) interval that covers their union.
	In a final step of the algorithm, both intervals will be assigned
	the color of their substitution.
	The remaining instance consists solely of pairs of events where two
	intervals start or two intervals end.
	Clearly, a balanced coloring has to assign opposite colors to the corresponding two
	intervals of such a pair, and any such assignment yields a balanced coloring.

	The remaining pairs of events induce a \emph{constraint graph}.
	Every vertex corresponds to an interval, and an edge is added
        between two vertices if there is a pair of events containing
        both startpoints or both endpoints. Finding a
	proper vertex two-coloring of this graph gives a balanced $2$-coloring.
	The constraint graph is bipartite: Each edge can
	be labeled by ``$\vdash$'' or ``$\dashv$'' if it corresponds to
	two start- or endpoints, respectively.
	Since each interval is incident to exactly two edges,
	any path must traverse $\vdash$- and $\dashv$-edges alternatingly.
	Therefore, every cycle must be of even length and hence the graph is bipartite.
	Thus, a proper vertex two-coloring of the constraint graph
	can be found in linear time by depth-first search.

  Sorting events takes $O(n \log n)$ time. Creation of
  the constraint graph and coloring it takes linear time.
\end{proof}

Note that if intervals are given already sorted, or interval endpoints are described by small integers, then the above algorithm can even find a balanced $2$-coloring in linear time. 

\subsection{Algorithms for $k$ Colors}
\label{sec:generalk}

In this section, we extend the results of the previous section to an arbitrary number of colors~$k$ and show that a balanced interval
$k$-coloring can be found in polynomial time.

A first polynomial time algorithm can be obtained using a construction by de Werra~\cite{Wer71}:
Start with an arbitrary coloring and find two colors~$i$ and~$j$ for which $\max_x |\col_i(x) - \col_j(x)|$ is maximal.
Use the algorithm from Section~\ref{sec:k2} to
find a balanced $2$-coloring of all intervals that currently have
color~$i$ or~$j$ and recolor them accordingly. Repeat until the coloring
is balanced. This algorithm has running time $O(n\log n + k^2 n)$,
because sorting intervals is needed only once for the above algorithm
for $2$ colors, and there are at most $\binom{k}{2}$ recolorings
necessary.

In the following, we present an alternative algorithm for $k$ colors, which is faster than $O(n\log n + k^2 n)$. We will first give an overview of the argument, and then a more formal description.

As in Section~\ref{sec:k2}, we assume w.~l.~o.~g. that all start- and endpoints
are pairwise disjoint. The idea is to scan the events in order, beginning with the smallest,
and to capture dependencies in $k$-tuples of intervals that indicate pairwise
different colors. That is, we reduce the \kbic~instance to an instance of
\textsc{Strong Hypergraph Coloring}, formally defined as follows.
\begin{center}
  \begin{minipage}{0.9\linewidth}
    \textsc{Strong Hypergraph Coloring}\\
    \textbf{Instance:} A ground set~$X$, a family~$\mathcal S$ of
    constraints $S_1, \dots, S_n \subseteq X$, and an integer~$k$.\\
    \textbf{Task:} Find a $k$-coloring~$\chi: X \to K$ with $\forall\ 1 \leq i \leq n: x, y \in S_i, x \neq y
    \Rightarrow \chi(x) \neq \chi(y)$.
  \end{minipage}
\end{center}
For example, in the special case that each block of $k$ consecutive events
consists only of start- or only of endpoints,
the constraints that the corresponding $k$ intervals have to be differently
colored will capture the whole solution.
As we will see below, also different interval nesting structures can be
captured by such constraints.

\textsc{Strong Hypergraph Coloring} is NP-hard in general~\cite{AH05}.
However, each interval will occur in at most two constraints,
corresponding to its start- and endpoint. Thus, we can further reduce
the \textsc{Strong Hypergraph Coloring} instance to \textsc{Edge
  Coloring}, where the goal is to color edges of a multigraph such
that the edges incident to each vertex are all differently colored.
More formally:
\begin{center}
  \begin{minipage}{0.9\linewidth}
    \textsc{Edge Coloring}\\
    \textbf{Instance:} A multigraph $G = (V, E)$ and an integer~$k$.\\
    \textbf{Task:} Find a $k$-coloring~$\chi: E \to K$ 
		with $\forall e, e' \in E, e \cap e' \neq \emptyset, e \neq e' \Rightarrow \chi(e) \neq \chi(e')$.
  \end{minipage}
\end{center}
It is easy to see that any instance $(X, \mathcal S, k)$ of
\textsc{Strong Hypergraph Coloring} where each element of~$X$ occurs
in at most two constraints can be reduced to \textsc{Edge Coloring} as
$(G = (\mathcal S, E), k)$, where for each $x \in X$ that occurs
in~$S_i$ and~$S_j$ with $i \neq j$, we add the edge~$\{S_i, S_j\}$
to~$E$. For our case, a constraint corresponds to a vertex, and an
interval corresponds to an edge that connects the two constraints it occurs in. An
edge coloring with $k$ colors will thus provide a balanced interval
$k$-coloring.

Clearly, an edge coloring of a multigraph with maximum degree
$\Delta$ needs at least $\Delta$ colors. Finding an edge coloring of
minimum size is NP-hard in general~\cite{Hol81}. However,
Kőnig~\cite{Kon16} showed that for bipartite multigraphs,
$\Delta$~colors in fact always suffice. Further, an edge coloring of a
bipartite multigraph with~$m$ edges can be found in $O(m \log \Delta)$
time~\cite{COS01}.
The multigraph we will construct has maximum degree~$k$
and is bipartite. Thus, a balanced interval $k$-coloring always exists and can
be found in polynomial time.

We now describe the construction of the constraints and give a more
rigorous description of the results. From a \kbic\ instance~$\mathcal I$,
we construct a \textsc{Strong Hypergraph Coloring} instance
$(\mathcal I, \mathcal S, k)$ over the ground set of the intervals.
The algorithm scans the set of events in order, beginning with the smallest.
It keeps a set of active events and adds constraints to~$\mathcal S$.
The set of active events will be cleared at each region where the number of intervals is
0 modulo~$k$. Thus, the active events always describe the change from a
situation where each color occurs the same number of times.

\begin{figure}
  \centering
  \includegraphics{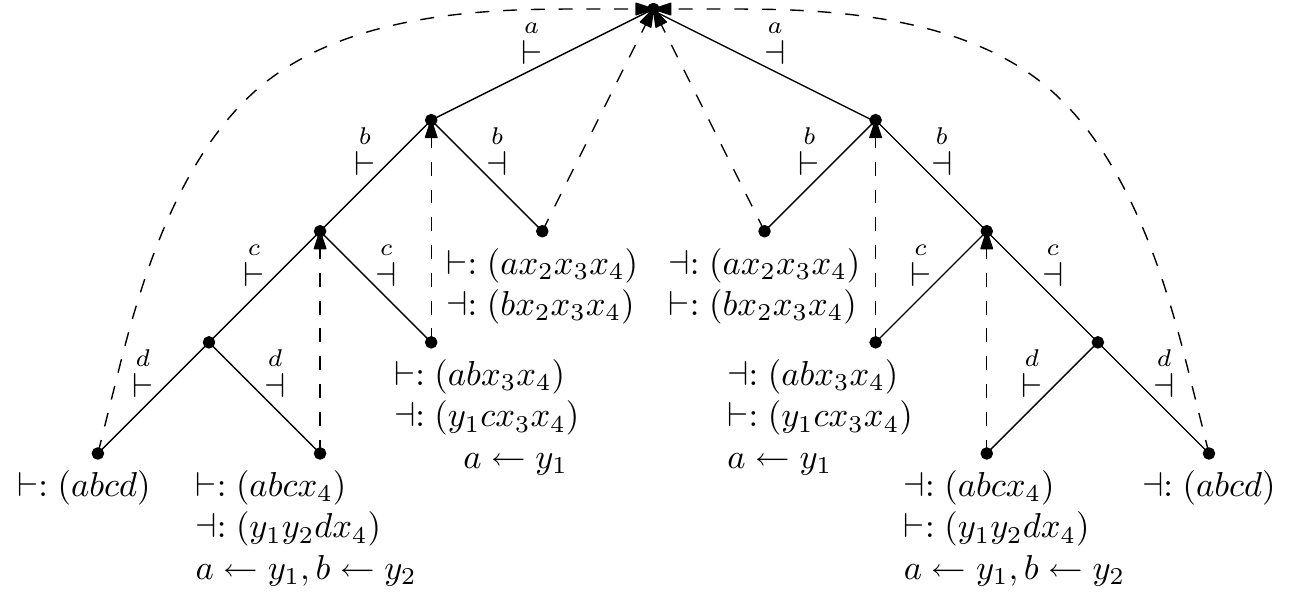}
  \caption{Tracking of active events for $k = 4$}
  \label{fig:dectree}
\end{figure}

The construction of the constraints can be visualized with a decision tree,
depicted for the example $k=4$ in Figure~\ref{fig:dectree}.
At the beginning, the set of active events is empty (which corresponds to the root
of the decision tree).
Whenever the set of events is empty, the algorithm branches into two
cases depending on the type of the next event.
Both branches are equivalent, with the roles of start- and endpoints interchanged.
Therefore, assume the next event is the start of an interval (depicted as $\stackrel{a}{\vdash}$ on the left
branch). It is added to the active set. This continues
until either $k$ startpoints of intervals~$I_1, \dots, I_k$ are added,
or an endpoint is encountered.
In the first case, the constraint
\begin{gather}
	(I_1,\dots, I_k) \label{cstr:3} 
\end{gather}
is constructed and
the set of active events is cleared (dashed arrow returning to the root).
In the second case, assume the startpoints of intervals
$I_1, \dots, I_j$ have been added and the endpoint of interval $I_{j+1}$ is encountered.
Then, the two constraints
\begin{gather}
  (I_1, \dots, I_j, x_{j+1}, \dots, x_k)              \label{cstr:1} \\ 
  (y_1, \dots, y_{j-1}, I_{j+1}, x_{j+1}, \dots, x_k) \label{cstr:2} 
\end{gather}
are constructed.
Here, $x_{j+1}, \dots, x_k$ and $y_1, \dots, y_{j-1}$ are new
``virtual'' intervals 
that have not been used in previous constraints. That is,
they do not correspond to actual intervals of the real line and only
serve as placeholders in the active set.
Furthermore, the startpoints of the intervals $I_1, \dots, I_j$
are replaced by the startpoints of $y_1, \dots, y_{j-1}$ in the active
set of events (indicated by the dashed arrows pointing one
level higher in the decision tree).

We now prove the correctness of the two chained reductions.

\begin{lemma}
  \label{lem:shc}
  A solution to the \textsc{Strong Hypergraph Coloring} instance
  $(\mathcal I, \mathcal S, k)$, constructed as described above,
	yields a balanced $k$-coloring for~$\mathcal I$.
\end{lemma}
\begin{proof}
	Recall that a region is an interval spanned by two consecutive events.
	The proof is by induction over all regions, in the order of events.
	At each region, we count the number of times each of the $k$ colors is used among
	the intervals containing the region. We will show that these counters
	differ by at most one, i.\,e., the coloring is balanced.
	Clearly, all counters are equal to zero before the first event.

	In particular, we show that in regions where the number of intervals is $0$~modulo~$k$,
	all counters are equal, and that between these regions the counters change by at most one
	and all in the same direction.
	We distinguish the same cases as in the construction, limiting the discussion
	to the case that the event first added to the empty set of
        active events is a startpoint.

  If $k$ startpoints of intervals~$I_1, \dots, I_k$ are encountered,
	there is a constraint of the form~\eqref{cstr:3}, which ensures
	that all of them have different
  colors. Therefore, at each startpoint, a different counter increases by one.
	Hence, the counters differ by at most one in all regions up to the
	startpoint of $I_k$, and are all equal
	in the region beginning with the startpoint of $I_k$.
	
  Consider that only $j< k$ startpoints of the intervals $I_1, \ldots, I_j$
	are encountered. Since these startpoints were added to the
	set of active events during the construction,
	the intervals $I_1,\ldots,I_j$ do all occur in one constraint.
	This constraint forces them to have different colors, and therefore
	the colors of $I_1,\ldots,I_j$ are exactly the colors with increased count.
	Now, we distinguish two subcases depending on the next event.
	If the next event is a startpoint of some interval, denoted by $I_{j+1}$, it will
	also be part of the same constraint. Hence, $I_{j+1}$ has a different color
	whose count is not yet increased.
	The second subcase is that the next event is the endpoint of some interval, denoted by $I_{j+1}$.
	The interval~$I_{j+1}$ is forced to have the same color as one of the
  intervals~$I_1, \dots, I_j$. This is because there is a constraint
	of the form~\eqref{cstr:1} that collects
  all other colors in variables $x_{j+1}, \dots, x_k$, which occur together
  with~$I_{j+1}$ in a constraint of the form~\eqref{cstr:2}.
	Thus, the previously increased counter for the color of~$I_{j+1}$ decreases again.
	Because of the constraint of the form~\eqref{cstr:1},
	the virtual intervals $y_1, \dots, y_{j-1}$ must
  have all of the colors of $I_1, \dots, I_j$ except for the color of
  $I_{j+1}$. Hence, the colors of $y_1,\ldots,y_{j-1}$ are exactly
	all remaining colors with increased count.
	Therefore, we are in the same situation as before encountering the
	endpoint of $I_{j+1}$. By repeating the above argument, the claim follows
	in this case.

	If the event first added is an endpoint, the argument is symmetrical.
\end{proof}

\begin{lemma}
  \label{lem:ec}
  A \textsc{Strong Hypergraph Coloring} instance $(\mathcal I,
  \mathcal S, k)$ constructed as described above can be reduced to a
  bipartite \textsc{Edge Coloring} instance.
\end{lemma}
\begin{proof}
	We need to show that each interval occurs in at most two constraints.
	Once this is proved, it is possible to build a multigraph with the constraints
	as vertices and edges between them if they share a common interval.
	Further, it has to be shown that this multigraph is bipartite.
	To show both parts at once, we color the constraints in $\mathcal S$ with the two colors
  $\vdash$ and $\dashv$. It then suffices to show that every interval
  can occur in at most one $\vdash$-constraint and in at most one
  $\dashv$-constraint.

	Consider Figure~\ref{fig:example_1} for an illustration of the constructed constraints and the respective bipartite \textsc{Edge Coloring} instance.
	\begin{figure}
		\centering
		\includegraphics[width=\textwidth]{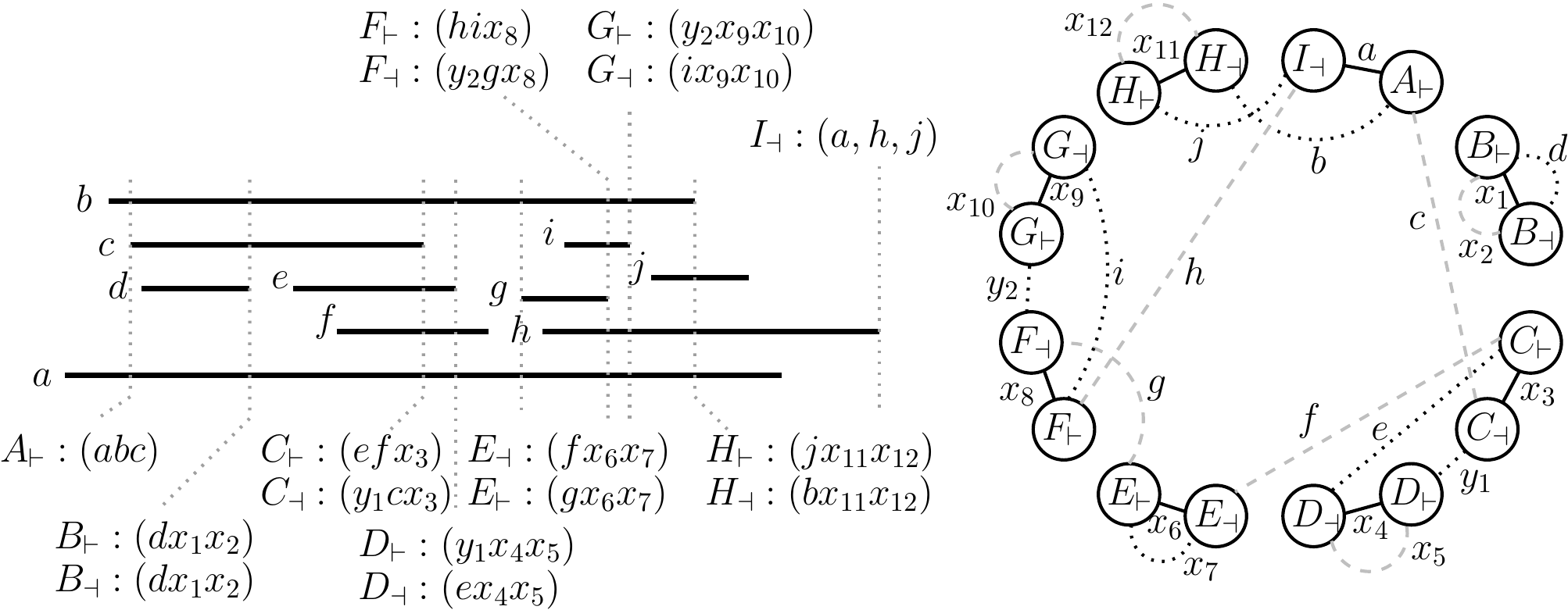}
		\caption{Complete example of the constraints and
			bipartite \textsc{Edge Coloring} instance for $k=3$ with a valid coloring.}
		\label{fig:example_1}
	\end{figure}

	We color a constraint
	with $\vdash$ if all involved events belonging to nonvirtual intervals are startpoints, and with $\dashv$ if all these events are endpoints (see Figure~\ref{fig:dectree}).
	All nonvirtual intervals therefore occur in exactly two constraints,
	constructed when the start- and endpoint get removed from the active set of events.
	A virtual $x$-interval always occurs in a pair of subsequent differently
  colored constraints, and is not used anywhere else.
	For the left branch of the decision tree, a virtual $y$-interval occurs first in a
  $\dashv$-constraint, and its startpoint is then added to the list of active events.
	Then, it will be used in a constraint of
  type either~\eqref{cstr:3} or~\eqref{cstr:1}, both
  of which are of type~$\vdash$.
	The argument is symmetrical for the right branch of the decision tree.
\end{proof}

\begin{theorem}
  \label{thm:generalk}
	Every set of $n$ intervals~$\mathcal I$ has a balanced $k$-coloring
	for any~$k\in\N$, and it can be found in~$O(n\log n +kn\log k)$ time.
\end{theorem}
\begin{proof}
	By Lemmas~\ref{lem:shc} and~\ref{lem:ec}, \kbic~can be reduced
	to \textsc{Edge Coloring} with fixed $k$ in a bipartite multigraph.
	The maximum degree of this multigraph is~$k$, since by construction no
	constraint has more than~$k$ elements.
	By Kőnig's theorem~\cite{Kon16} existence follows.

  To be able to process the events, they have to be sorted in~$O(n
  \log n)$ time. We have~$O(kn)$ virtual intervals and thus~$O(kn)$
  edges in the \textsc{Edge Coloring} instance. Finding an edge
  $k$-coloring for this multigraph with maximum degree~$k$ can be done in
  $O(kn \log k)$ time~\cite{COS01}.
\end{proof}

Note that the \textsc{Edge Coloring} algorithm by Cole, Ost, and
Schirra~\cite{COS01} uses quite involved data structures. In practice,
it might be preferable to use the much simpler algorithm by
Alon~\cite{Alo03} running in $O(m \log m)$ time for an $m$-edge graph,
which gives a worst-case bound of $O(kn \log n)$.
An implementation of our algorithm in Python using a simple edge
coloring algorithm based on augmenting paths can be found at
\url{http://www2.informatik.hu-berlin.de/~hueffner/intcol.py}.

For an extension, recall that a matrix has the consecutive-ones
property if there is a permutation of its columns such that all
$1$-entries appear consecutively in every row. Such a permutation can
be found in linear time by the PQ-algorithm~\cite{BL75}. Given such a
matrix, it is straightforward to construct an instance of \kbic.

\begin{theorem}
For any hypergraph $H$ with an $n \times m$ incidence matrix having
the consecutive ones property, a balanced $k$-coloring can be found in
$O(nm + k n \log k)$ time.
\end{theorem}

\subsection{Arcs of a Circle}
In a periodic setting, the tasks $\mathcal{I}$ might be better
described by a set of arcs of a circle rather than a set of
intervals. In this case, there are instances that require an imbalance
of two (e.\,g., three arcs that intersect exactly pairwise and
$k=2$). We show that two is also an upper bound and a coloring with
maximal imbalance two can be found in polynomial time.

\begin{theorem}
  \label{thm:arcs}
  The maximal imbalance for arcs of a circle is two, and finding a
  coloring with imbalance at most two can be done in~$O(n \log n+kn\log k)$
  time.
\end{theorem}
\begin{proof}
	Define a point on the circle, called zero, and consider counterclockwise orientation.
	We build an instance of \kbic\ by ``unfolding'' the circle at zero in the following way.
	Consider only arcs that do not span the full circle.
	Map all such arcs not containing zero to intervals of same length
	at the same distance right of zero on the real line.
	Map the arcs containing zero to intervals of same length such that
	the positive part of the interval has the same length as the part of the arc in counterclockwise direction from zero.
	Finally, map the arcs containing the full circle to intervals spanning all of the instance constructed so far.
	Use the above algorithm to obtain a coloring of the intervals with
	imbalance at most one at every point.
	By reversing the mapping, the obtained coloring of the arcs has
	imbalance at most two (each point on the circle is mapped to at most two points
	of the real line).
%
\end{proof}

\subsection{Online Algorithms}

In load balancing problems, it is often more realistic to assume an online
scenario, where not all information is known in advance, but is 
rather arriving piece-by-piece, and irrevocable decisions have to be made
immediately. In our setting, this means that intervals arrive in order
of their startpoint, including the information of their endpoint, and
a color has to be assigned to them immediately.

The problem of finding a proper coloring (i.\,e., a coloring where no
two intersecting intervals have the same color) of intervals in an
online setting has found considerable
interest~\cite{KT81,Eps08}. 
In these works, the objective is to use a minimum number of colors.
In contrast, we consider a fixed number of colors and the minimization of
the imbalance. We show that in contrast to the offline scenario, here
the imbalance can become arbitrarily large.

\begin{theorem}\label{onlinethm}
  In online \kbic, the imbalance is unbounded.
\end{theorem}
\begin{proof}
	We first consider the case $k = 2$ with colors ``$+1$'' and ``$-1$''. Denote the signed
	imbalance $\simb(x)$ to be the sum of the colors of the intervals containing $x$.
	Note that $\imb(x) = |\simb(x)|$.

	In the following, we outline how a sequence of intervals can be constructed
	such that no online algorithm can yield a bounded imbalance.
	Initially, $\simb\equiv 0$.
	Set $L=[0,1]$ and $R=[2,3]$.
	Let $L_\ell$, $R_\ell$, $L_r$, and $R_r$ denote the start- and endpoints
	of the current $L$ and $R$, respectively.
	Repeat the following steps.
  \begin{itemize}
  \item Present the interval $[(L_\ell + L_r)/2, (R_\ell + R_r) / 2]$ to the
    online algorithm.
  \item If it chooses color~$+1$, set $R \leftarrow [R_\ell, (R_\ell + R_r)
    / 2]$, else $R \leftarrow [(R_\ell + R_r) / 2, R_r]$.
  \item Set $L \leftarrow [(L_\ell + L_r)/2, L_r]$.
  \end{itemize}
  This sequence of intervals is legal, since the startpoints increase strictly monotonously.
	In each repetition, if the algorithm chooses color $+1$, the
	signed imbalances in $L$ and $R$ increase by one. If the algorithm chooses~$-1$,
	the signed imbalance decreases by one in $L$ and remains unchanged in $R$, i.\,e.,
	the difference of the signed imbalance in $L$ and $R$ increases.
	Therefore, the signed imbalance diverges in $L$ or $R$.
	Since the imbalance is the absolute value of the signed imbalance, it becomes unbounded.

	The construction easily generalizes to $k > 2$ colors. We only track two arbitrary
	colors, and whenever the algorithm assigns an untracked color to
	an interval, we present the same interval (with a slightly increased startpoint) again, forcing it to
	eventually assign a tracked color or to produce unbounded imbalance.
\end{proof}

\section{Hardness of Generalizations}
\label{section:extensions}

We consider several generalizations of \kbic\ and show that they are
NP-hard. Note that the hardness results of Biedl et al.~\cite{BCC02}
do not apply to the problems we consider here.

\subparagraph{$d$-Dimensional Boxes.}

Gyárfás and Lehel~\cite{GL85} suggest to examine $d$-dimensional boxes
as generalizations of intervals for coloring problems. The problem
\dkbbc\ has as input an integer $k$ and a set $\mathcal I=\{I_1,\dots,
I_n\}$ of $n$ $d$-dimensional boxes $I_i= ([\ell_{i,1},
  r_{i,1}],[\ell_{i,2}, r_{i,2}],\dots,[\ell_{i,d}, r_{i,d}])$ for $1
\le i \le n$.

For every point $x=(x_1, \dots, x_d)$, let $S(x)$ be the set of boxes that include $x$, i.\,e., $S(x)$ contains all the elements $I_i$ such that $\ell_{i,j}\le x_j \le r_{i,j}$ for all $1 \le j \le d$. For a coloring $\chi : \mathcal{I} \rightarrow K$, a color~$i$, and a point~$x$, let~$\col_i(x)$ be the number of boxes in $S(x)$ of color $i$. With the analog definition of imbalance $\imb(\chi)$ and balance for $d$-dimensional boxes, the problem statement becomes:

\begin{center}%
  \begin{minipage}{0.9\linewidth}%
    \dkbbc\\
    \textbf{Instance:} A set $\mathcal{I}$ of $d$-dimensional boxes.\\
    \textbf{Task:} Find a $k$-coloring $\chi$ with minimal $\imb(\chi)$.
  \end{minipage}%
\end{center}%

First note that, unlike for the case $d=1$, a balanced coloring may
not exist: already for three rectangles, some instances require
imbalance two. Hence, we also have a related decision problem:

\begin{center}%
  \begin{minipage}{0.9\linewidth}%
    \ibbc\\
    \textbf{Instance:} A set $\mathcal{I}$ of $d$-dimensional boxes.\\
    \textbf{Question:} Is there a balanced $k$-coloring $\chi$?
  \end{minipage}%
\end{center}%

We show that for all $d\ge 2$ and $k\ge 2$, it is NP-complete to decide \ibbc. This clearly implies NP-hardness of \dkbbc.

\begin{theorem}
\ibbc\ is NP-complete for any $d\ge2$ and any $k\ge 2$.
\label{thm:npc}
\end{theorem}

We will reduce from \textsc{Not-All-Equal 3SAT
  (NAE-3SAT)}~\cite{Schaef78}. Note that the classic definition of
NAE-3SAT~\cite{GareyJ79} allows negated variables. However, this is
not needed to make the problem NP-complete~\cite{Schaef78}. Thus, in
the sequel, we will assume that all variables occur only
non-negatedly.

\begin{center}%
  \begin{minipage}{0.9\linewidth}%
    \textsc{Not-All-Equal 3SAT (NAE-3SAT)}\\
    \textbf{Instance:} A Boolean formula with clauses $C_1, \dots, C_m$, each having at most $3$ variables.\\
    \textbf{Question:} Is there a truth assignment such that in every clause, not all variables have the same value?
  \end{minipage}%
\end{center}%

We first consider $k = 2$ and then generalize to arbitrary $k$. We present the gadgets of the reduction, then show how they are combined together, and conclude by proving correctness. 

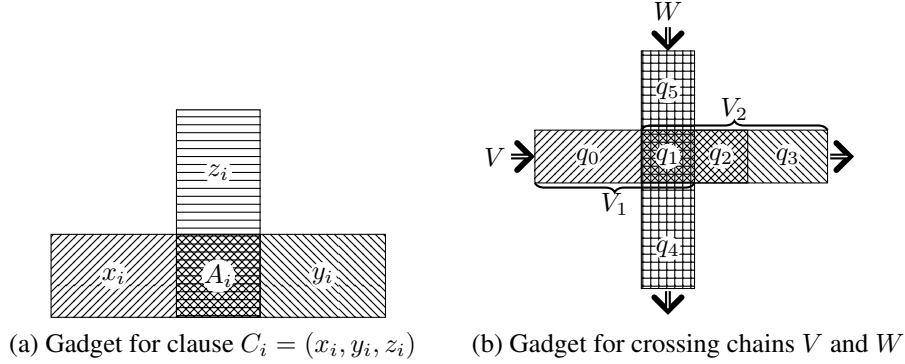
\begin{figure}
\centering
     \begin{tikzpicture}[scale=0.55]
       \draw[pattern=north east lines] (0,0) rectangle (5,2);
       \draw[pattern=north west lines] (3,0) rectangle (8,2);
       \draw[pattern=horizontal lines] (3,0) rectangle (5,5);
       \draw[color=white,fill=white] (1.5,1) circle (9pt);
       \draw[color=white,fill=white] (6.5,1) circle (9pt);
       \draw[color=white,fill=white] (4,3.5) circle (9pt);
       \draw[color=white,fill=white] (4,1) circle (11pt);
       \node at (1.5,1) {$x_i$};
       \node at (6.5,1) {$y_i$};
       \node at (4,3.5) {$z_i$};
       \node at (4,1) {$A_i$};
     \end{tikzpicture}
	\hspace{1cm}
  \begin{tikzpicture}[scale=0.35]
    \draw[pattern=north east lines] (0,0) rectangle (8,2);
    \draw[pattern=north west lines] (11,0) rectangle (4,2);
    \draw[pattern=grid] (4,5) rectangle (6,-4);
    \draw[color=white, fill=white] (5,1) circle (13pt);
    \draw[color=white, fill=white] (7,1) circle (13pt);
    \draw[color=white, fill=white] (9.5,1) circle (13pt);
    \draw[color=white, fill=white] (2,1) circle (13pt);
    \draw[color=white, fill=white] (5,-2.5) circle (13pt);
    \draw[color=white, fill=white] (5,3.5) circle (13pt);
    \node at (5, 1) {$q_1$};
    \node at (7, 1) {$q_2$};
    \node at (9.5, 1) {$q_3$};
    \node at (2, 1) {$q_0$};
    \node at (5, -2.5) {$q_4$};
    \node at (5, 3.5) {$q_5$};
    \draw [->,double,shorten <=1pt,>=angle 90,thick](-1,1) -- (0,1);
    \draw [->,double,shorten <=1pt,>=angle 90,thick](11,1) -- (12,1);
    \node at (-1.5,1) {$V$ };
    \draw [->,double,shorten <=1pt,>=angle 90,thick](5,6) -- (5,5);
    \draw [->,double,shorten <=1pt,>=angle 90,thick](5,-4) -- (5,-5);
    \node at (5,6.5) {$W$};
    \draw[decorate,decoration={brace, amplitude=3.5pt},thick] (6,0) to
	node[midway,below] (bracket) {$V_1$}	(0,0);
    \draw[decorate,decoration={brace,amplitude=3.5pt},thick] (4,2) to
	node[midway,above] (bracket) {$V_2$}	(11,2);
  \end{tikzpicture}

\begin{minipage}{5.5cm}
\centering
(a) Gadget for clause $C_i = (x_i, y_i, z_i)$
\end{minipage}
\hspace{0.5cm}
\begin{minipage}{5.7cm}
\centering
(b) Gadget for crossing chains $V$ and $W$
\end{minipage}
\caption{The gadgets of the reduction.}
\label{fig:clause_crossing}
\end{figure}

\begin{figure}
  \centering

  \begin{tikzpicture}[scale=0.3]
   \draw[fill=gray, opacity=0.5] (0,3) rectangle (5,5);
   \draw[fill=gray, opacity=0.5] (3,3) rectangle (8,5);
   \draw[fill=gray, opacity=0.5] (3,3) rectangle (5,8);
   \node at (1.5,4) {$x_1$};
   \node at (6.5,4) {$x_3$};
   \node at (4,6.5) {$x_2$};

   \draw[fill=gray, opacity=0.5] (9,3) rectangle (14,5);
   \draw[fill=gray, opacity=0.5] (12,3) rectangle (17,5);
   \draw[fill=gray, opacity=0.5] (12,3) rectangle (14,8);
   \node at (10.5,4) {$x_3$};
   \node at (15.5,4) {$x_5$};
   \node at (13,6.5) {$x_4$};

   \draw[fill=gray, opacity=0.5] (18,3) rectangle (23,5);
   \draw[fill=gray, opacity=0.5] (21,3) rectangle (26,5);
   \draw[fill=gray, opacity=0.5] (21,3) rectangle (23,8);
   \node at (19.5,4) {$x_1$};
   \node at (24.5,4) {$x_6$};
   \node at (22,6.5) {$x_5$};

   \draw[fill=gray, opacity=0.5] (0,16) rectangle (4,18);
   \node at (2,17) {$x_1$};

   \draw[fill=gray, opacity=0.5] (4.4,16) rectangle (8.4,18);
   \node at (6.4,17) {$x_2$};

   \draw[fill=gray, opacity=0.5] (8.8,16) rectangle (12.8,18);
   \node at (10.8,17) {$x_3$};

   \draw[fill=gray, opacity=0.5] (13.2,16) rectangle (17.2,18);
   \node at (15.2,17) {$x_4$};

   \draw[fill=gray, opacity=0.5] (17.6,16) rectangle (21.6,18);
   \node at (19.6,17) {$x_5$};

   \draw[fill=gray, opacity=0.5] (22,16) rectangle (26,18);
   \node at (24,17) {$x_6$};

   \draw[fill=gray, opacity=0.5] (25,16.5) rectangle (25.5,4.5);
   \draw[fill=gray, opacity=0.5] (21,16.5) rectangle (21.5,7.5);
   \draw[fill=gray, opacity=0.5] (1,16.5) rectangle (1.5,10);
   \draw[fill=gray, opacity=0.5] (1,10.5) rectangle (5.4,10);
   \draw[fill=gray, opacity=0.5] (4.4,10.5) rectangle (8.4,10);
   \draw[fill=gray, opacity=0.5] (7.5,10.5) rectangle (10.5,10);
   \draw[fill=gray, opacity=0.5] (9.5,10.5) rectangle (14.5,10);
   \draw[fill=gray, opacity=0.5] (13.5,10.5) rectangle (17.5,10);
   \draw[fill=gray, opacity=0.5] (16.5,10.5) rectangle (19.5,10);
   \draw[fill=gray, opacity=0.5] (19,10.5) rectangle (19.5,7);
   \draw[fill=gray, opacity=0.5] (19,4.5) rectangle (19.5,7.5);

   \draw[fill=gray, opacity=0.5] (0,16.5) rectangle (0.5,4.5);
   \draw[fill=gray, opacity=0.5] (4.4,16.5) rectangle (4.9,7.5);
   \draw[fill=gray, opacity=0.5] (8.8,16.5) rectangle (9.3,14);
   \draw[fill=gray, opacity=0.5] (9.3,14) rectangle (7.5,14.5);
   \draw[fill=gray, opacity=0.5] (7.5,14.5) rectangle (8,4.5);

   
   \draw[fill=gray, opacity=0.5] (9.5,16.5) rectangle (10,4.5);
   \draw[fill=gray, opacity=0.5] (13.5,16.5) rectangle (14,7.5);
   \draw[fill=gray, opacity=0.5] (18.6,16.5) rectangle (18.1,14);
   \draw[fill=gray, opacity=0.5] (16.5,14) rectangle (18.6,14.5);
   \draw[fill=gray, opacity=0.5] (16.5,14.5) rectangle (17,4.5);

   \end{tikzpicture}
  \caption{Example for \textsc{NAE-3SAT} instance $(x_1,x_2,x_3),(x_3,x_4,x_5),(x_1,x_5,x_6)$.}
  \label{fig:example}
\end{figure}
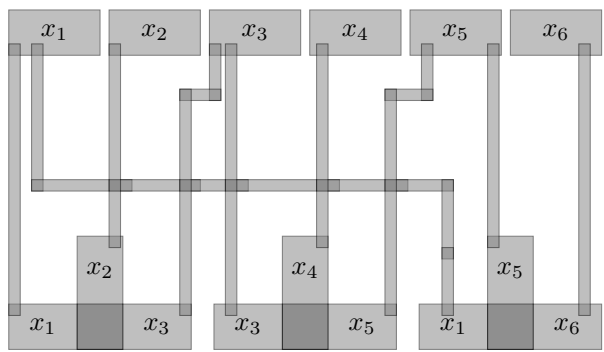

For each clause $C_i=(x_i,y_i,z_i)$, we construct a \emph{clause
  gadget} comprised of three rectangles (see
Figure~\ref{fig:clause_crossing}a). Note that all three rectangles
overlap in region $A_i$, and only there. Then we also construct a
separate rectangle $r_j$ for every variable. Finally, we connect each
$r_j$ to all rectangles that appear in a clause gadget, and correspond
to the same variable as $r_j$. We do this by a chain with odd number
of rectangles. This ensures that in any balanced $2$-coloring, $r_j$
and the corresponding rectangle in the clause gadget have the same
color. If two chains need to cross, we introduce a \emph{crossing
  gadget} as seen in Figure~\ref{fig:clause_crossing}b. Three
rectangles are relevant for the crossing of two chains $V$ and $W$.
The first is $V_1$ and contains areas $q_0$, $q_1$, and $q_2$, the
second is $V_2$, containing $q_1$, $q_2$, and $q_3$. Both $V_1$ and
$V_2$ belong to chain $V$. The last rectangle contains areas $q_1,q_4$
and $q_5$ and belongs to chain $W$. Note that the crossing does not
induce any dependencies on the colorings between chains $V$ and
$W$. See Figure~\ref{fig:example} for a construction of an instance
for \dbbc{2}{2}.

Observe that the above construction only requires a number of rectangles polynomial in the size of the NAE-3SAT instance. 

\begin{lemma}
\dbbc{d}{2}\ is NP-complete for any $d\ge2$.
\end{lemma}

\begin{proof}
The problem is in NP, since feasibility of a color assignment can be
checked in polynomial time. For NP-hardness, we show that a NAE-3SAT
instance is satisfiable if and only if the answer to the corresponding
\dbbc{2}{2}\ instance is ``yes''. This also implies NP-completeness
for every $d\ge 2$ by taking intervals of length $0$ in higher
dimensions.

($\Rightarrow$) Assume that there is a satisfying assignment of the NAE-3SAT instance. Then, color the rectangles $r_j$ according to the truth values of their corresponding variables. This coloring can be easily extended to all the rectangles by alternatively coloring rectangles along a chain (and crossings) starting from each $r_j$ and ending at a clause gadget. It remains to show that $\imb(x) \le 1$ holds for every point $x \in A_i$ for all $1\le i \le m$. Consider $A_i$ corresponding to clause $C_i=(x_i,y_i,z_i)$. The three rectangles that intersect at $A_i$ have the colors corresponding to the truth values of their variables $x_i,y_i,$ and $z_i$ in the solution of NAE-3SAT. Since the three variables do not have all the same truth value, the three rectangles cannot have all the same color, and $\imb(x) \le 1$.

($\Leftarrow$) Assume that we have a balanced 2-coloring for the constructed \dbbc{2}{2}\ instance. Consider only the clause gadgets. We have already observed that rectangles that correspond to the same variable and appear in clause gadgets must have the same color. We can assign the truth values of the variables according to the colors in the corresponding rectangles. Since in no $A_i$ all three rectangles have the same color, in no $C_i$ all three variables have the same truth value, yielding a feasible solution for NAE-3SAT.
\end{proof}

\begin{proof}[Proof of Theorem~\ref{thm:npc}]
First apply the construction for \dbbc{2}{2}\ and call its rectangles \emph{reduction rectangles}. Then add $k - 2$ additional rectangles that fully contain the construction and all intersect at least in one point outside the construction; these are called \emph{cover rectangles}. By the latter property, cover rectangles must have distinct colors in any balanced coloring. Observe that each reduction rectangle contains some point that does not intersect with other reduction rectangles but only with all the cover rectangles. This implies that the   reduction rectangles have available only the two colors not used by the cover rectangles. We conclude that the problem of $k$-coloring the constructed instance is equivalent to the problem of $2$-coloring only the reduction rectangles.
\end{proof}

\subparagraph{Further Generalizations.}
The weighted version, where intervals have weights and the weighted
imbalance is to be minimized, is NP-complete by reduction from
\textsc{Partition}.
Furthermore, the variant with multiple intervals~\cite{GL85} is NP-complete by reduction from \textsc{NAE-3SAT}. Both hardness results generalize to higher dimensions. The proofs can be found in Appendix~\ref{sec:appendix}.

\section{Open Questions}

\begin{itemize}
\item We have given a polynomial time algorithm for $k$-coloring
  hypergraphs with the consecutive-ones property, i.\,e., a special case
  of a totally unimodular incidence matrix. It would be interesting to
  generalize to arbitrary totally unimodular incidence matrices.
\item For arcs of a circle, we have shown how to find a coloring with imbalance at most two in polynomial time, but it is not clear how to find an optimal one.
\item It remains open how large the imbalance can become for $d$-dimensional boxes, and whether we can find polynomial-time approximations for it. We were not able to find an instance requiring an imbalance greater than $2$ for the $2$-dimensional case. 
\end{itemize}

\bibliographystyle{abbrv}
\bibliography{intcol}

\newpage
\appendix
\section{Appendix}
\label{sec:appendix}

We consider further generalizations that can be shown to be NP-complete.
\subsection{Weighted Version}

We consider the variant where each interval $I \in \mathcal{I}$ has a weight $w_I \in \N$,
and we want to $k$-color $\mathcal{I}$ as evenly as possible. For a
coloring $\chi : \mathcal{I} \rightarrow K$, let $w_i(x)$ denote the
total weight of the intervals containing $x$ that have color $i$.
Then the weighed imbalance is $\imb(x) = \max_{i,j \in K} |w_i(x) -
w_j(x)|$, which we seek to minimize.

Even for $k = 2$, it is NP-complete to decide if the minimum weighted imbalance
of $\mathcal{I}$ is zero. This also implies NP-hardness of the
minimization problem. We will reduce from {\sc Partition}
\cite{GareyJ79}: Given a set $A=\{a_1,a_2,\dots, a_n\}$ and a cost
function $s: A \rightarrow \N$, is there a subset $A'\subseteq A$ such
that $\sum_{a\in A'}s(a) = \sum_{a\in A\setminus A'}s(a)$? The
reduction is straightforward: For any given instance of {\sc
  Partition} of size $n$, construct $n$ identical intervals $[\ell,r]$
for arbitrary $\ell<r$, with each interval $I_i$, for $1\le i\le n$,
having weight $s(a_i)$. This instance has an exactly balanced color
assignment if and only if the answer to the corresponding {\sc
  Partition} instance is ``yes''.

Note that the above reduction is for $1$-dimensional problems, but
implies the NP-completeness of higher-dimensional problems, too.

\subsection{Multiple intervals}

Another generalization suggested by Gyárfás and Lehel~\cite{GL85} is
\emph{multiple intervals}, where specific subsets of non-intersecting
intervals must receive the same color. This variant is also
NP-complete. Given a NAE-3SAT instance where no negations are allowed,
for every clause $C_i$ construct three intervals corresponding to the
three variables of the clause. All these three intervals have the same
startpoints $\ell_i$ and endpoints $r_i$ and $\ell_i>r_{i-1}$. Finally, for
every variable, pack all the corresponding constructed intervals into
a subset that enforces that they receive the same color.
This is legal, since different intervals for the same variable are
disjoint, and  it can be easily seen that the
multiple intervals instance has a balanced coloring if and only if the
corresponding NAE-3SAT instance has a satisfying assignment.

Again, the hardness generalizes to higher dimensions.

\end{document}